\documentclass[conference]{IEEEtran}
\IEEEoverridecommandlockouts
\usepackage{spconf,amsmath,graphicx}
\usepackage[english]{babel}
\usepackage{soul}
\usepackage{hyperref}
\usepackage[capitalize]{cleveref}
\usepackage{tabularray}
\usepackage[dvipsnames]{xcolor}

\usepackage{amssymb,bm}
\usepackage{textcomp}
\usepackage{booktabs}
\usepackage[textfont=it,tableposition=top]{caption}
\usepackage{siunitx}
\usepackage{xspace}
\usepackage{url}
\usepackage{lipsum}  
\usepackage{tipa}
\usepackage{enumitem}
\usepackage{cite}
\usepackage{algorithmic}
\usepackage{dblfloatfix}
\def\BibTeX{{\rm B\kern-.05em{\sc i\kern-.025em b}\kern-.08em
    T\kern-.1667em\lower.7ex\hbox{E}\kern-.125emX}}

\usepackage[activate]{microtype}
\newcommand{\eg}{e.\,g., }
\newcommand{\ie}{i.\,e., }

\newcommand{\evits}{ParaEVITS }

\def\ninept{\def\baselinestretch{.95}\let\normalsize\small\normalsize}

\begin{document}
\title{Enhancing Emotional Text-to-Speech Controllability with Natural Language Guidance through Contrastive Learning and Diffusion Models
}
\ninept

\name{Xin Jing$^{1,2}$, Kun Zhou$^{4}$, Andreas Triantafyllopoulos$^1$, Bj\"orn W.\ Schuller$^{1,3}$}
\address{
  $^1$CHI – Chair of Health Informatics, Technical University of Munich, Munich, Germany\\
  $^2$Chair of Embedded Intelligence for Health Care \& Wellbeing, University of Augsburg, Germany\\
  $^3$GLAM -- Group on Language, Audio, \& Music, Imperial College London, UK\\
  $^4$Alibaba Group, Singapore\\[0.5em]
  \texttt{xin.jing@tum.de}}

\maketitle
\begin{abstract}

While current emotional text-to-speech (TTS) systems can generate highly intelligible emotional speech, achieving fine control over emotion rendering of the output speech still remains a significant challenge.
In this paper, we introduce ParaEVITS, a novel emotional TTS framework that leverages the compositionality of natural language to enhance control over emotional rendering. By incorporating a text-audio encoder inspired by ParaCLAP, a contrastive language-audio pretraining~(CLAP) model for computational paralinguistics, the diffusion model is trained to generate emotional embeddings based on textual emotional style descriptions.
Our framework first trains on reference audio using the audio encoder, then fine-tunes a diffusion model to process textual inputs from ParaCLAP's text encoder.
During inference, speech attributes such as pitch, jitter, and loudness are manipulated using only textual conditioning. Our experiments demonstrate that \evits effectively control emotion rendering without compromising speech quality. Speech demos are publicly available\footnote{https://synaudio.github.io/syn\_demo}.

\end{abstract}
\section{Introduction}
\label{sec:intro}
Emotional text-to-speech (TTS) has become a crucial technique in human-computer interaction, enabling virtual agents to generate natural-sounding speech that conveys emotions aligned with the conversational context \cite{Andreas24-EAS, Triantafyllopoulos24-BDL, Inoue24-HED}.
While both the speech quality and emotional intelligibility of emotional TTS have improved through the use of deep learning in recent years, \emph{controllability} remains a key open challenge \cite{Trian23-AOO,Chandra24-ESS3, Zhou23-EIA}.
To capture emotional information, previous emotion TTS systems typically relied on either a fixed set of emotion labels (i.e., ``one-hot'' encoding) or the transfer of emotions from reference speech~\cite{Huang23-PIP, Kang23-ZSZ, Kang23-GAA}.
The former often results in stereotypical emotional patterns, while the latter offers limited control over emotional rendering, as speech signals encode multiple entangled attributes \cite{Zhou21-VGF, Zhou22-EVC}.
However, emotional TTS systems require more nuanced and fine-grained emotional representations %\textcolor{red}{What does "a higher degress of expressivity" mean?} 
than one-hot encoding \cite{Inoue24-HEP, Andreas24-EAS}. Moreover, mimicking an emotional style from a reference utterance significantly limits application usability, such as the challenge of selecting appropriate references for a conversational assistant \cite{Bhardwaj24-CAA,Chawla23-TEA}.

One approach to improving the controllability of TTS models is by using natural language to guide the synthesis process \cite{Liu24-A2L, Liu23-PCS, Guo23-PCT, Bott24-CEI}.
Recent advancements in prompt engineering and multimodal pre-training, initially developed for image generation\cite{Zhang23-TDM, Croitoru23-DMI}, have been successfully extended to audio generation \cite{Liu24-A2L, Huang23-MTG} and speech synthesis \cite{Liu23-PCS, Guo23-PCT, Bott24-CEI}.
For instance, PromptStyle  \cite{Liu23-PCS} and InstructTTS \cite{Yang24-IME} integrate a cross-modal style encoder into a TTS system to enable controllable style transfer using natural language descriptions.
Additionally, PromptTTS  \cite{Guo23-PCT} synthesizes speech by taking prompts that specify both style and content. However, these systems heavily rely on manually annotated captions. Even though PromptTTS utilizes LLMs to assist in caption engineering, it still requires significant human intervention for accurate labeling. PromptTTS 2 \cite{Leng23-PDA} attempts to address this challenge by automatically labeling voice attributes such as gender and speed from audio and generating descriptive prompts accordingly. Despite this improvement, the lack of detailed descriptions limits the level of fine-grained control over the generated speech.

Recently, ParaCLAP \cite{Jing24-PTA} was proposed to generate rich emotional style descriptions, showing promising results in speech recognition tasks. It couples Computational Paralinguistics (CP) tasks, which focus on analyzing non-verbal vocal cues such as tone and pitch, with natural language processing through Contrastive Language-Audio Pretraining (CLAP) to more effectively characterize the emotional and expressive nuances in human speech. This approach enhances the automatic generation of speech captions by effectively leveraging CP attributes.
In this paper, we propose a contrastive learning method based on ParaCLAP \cite{Jing24-PTA} to condition a diffusion model, emphasizing stylistic attributes that are overlooked by previous work. Specifically, inspired by the literature on CP~ \cite{Schuller14-CPE}, we rely on prompts explicitly designed to capture paralinguistic attributes.
\begin{figure*}[ht]
    \centering
    \vspace{-5mm}
    \includegraphics[width=0.82\textwidth]{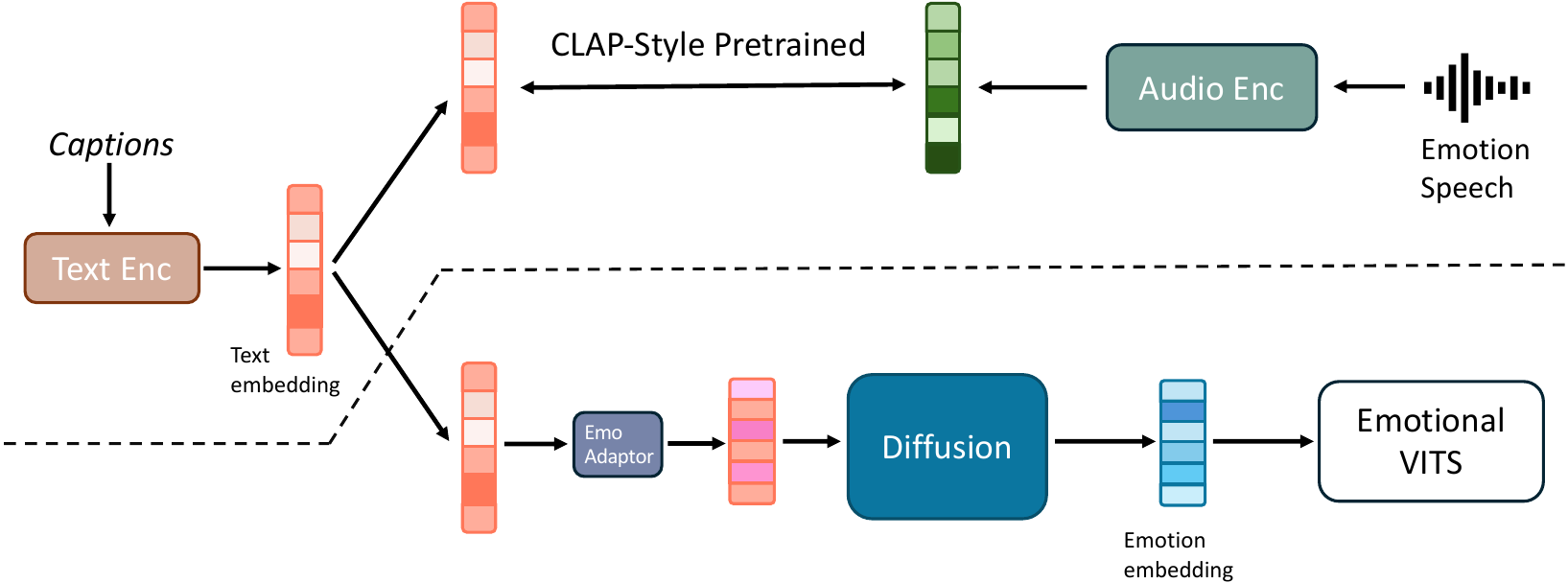}
    \caption{A high-level overview of the \evits concept. Above the dotted line is the training process of the ParaCLAP-NP, through which we learn an aligned representation for emotion and captions. Below the dotted line is emotional TTS generation process: a ParaCLAP-NP text embedding is first processed by the diffusion prior to produce the ParaCLAP emotion embedding, and then, Emo Adaptor will map the emotion embedding into a smaller size, which will be used as the emotion input of the emotional VITS to synthesis a final waveform. The ParaCLAP-NP and Emo Adaptor are frozen during the training of the diffusion model.}
    \label{fig:overview}
\end{figure*}
CP-based prompts can address a wide array of phenomena, including classic emotions, personality, likability, sincerity, deception, and health-related tasks\cite{Schuller14-CPE, Zhang13-CSI}. ParaCLAP identified six key variables that incorporate both low-level and high-level emotion features, strongly linked to emotional styles: \textbf{pitch, pitch variation, sound level, jitter, shimmer}, and \textbf{emotion labels}.
These variables allow for precise control over various audio attributes, effectively guiding our TTS system in synthesizing the final emotional speech. Our novel contributions are summarized as follows:
\begin{itemize}
    \item We adopt a Computational Paralinguistics (CP)-based prompt engineering approach that bridges the gap between low-level acoustic features and high-level emotion descriptions. Our method enables precise control over the emotional rendering of synthesized speech, allowing users to effectively control the fine-grained details of the system's emotional output.
    \item We demonstrate the diffusion process's capability to generate effective natural language emotion embeddings. These embeddings can guide the TTS framework to produce speech that aligns with the natural language prompts provided. 
    \item We propose ParaEVITS, moving beyond traditional emotional TTS methods that depend on pre-defined emotional categories or reference speech. Instead, we bridge CP features and natural language captions to more effectively model and control emotion in synthesized speech;
\end{itemize}

The rest of this paper is organized as follows:
Our proposed \evits is introduced in \cref{sec:sys}.
\cref{sec:exp} presents our caption generation and experiments details. Our results and analysis are discussed in \cref{sec:res}. 
Finally, the conclusion and future work are discussed in \cref{sec:con}.

\section{Proposed Method}
\label{sec:sys}
\subsection{Model Architecture}
Each sample in our training data comprises $(x, y, z, text)$ -- a tuple  of audio $x$, emotion embedding $y$, corresponding captions $z$, and transcriptions $text$. More details will be discussed in 
Section \ref{ssec:dataset}. As illustrated in \cref{fig:overview}, we design our generative stack in \evits to produce emotional speech \textbf{guided only by captions} using two main components:

\begin{itemize}
   \item A diffusion model as prior $P(y_i|z)$ produces ParaCLAP-style emotion embeddings $y_i$ conditioned on ParaCLAP-style caption embeddings $z$;
    \item A TTS model as decoder $P(x|y_i, text)$ synthesizes emotional speech $x$ conditioned on the ParaCLAP-style emotion embeddings $y_i$ and content embedding $text$.
\end{itemize}

The prior allows us to learn a generative model of the emotion embeddings themselves, while the decoder allows us to generate the emotional speech given emotion embeddings and content embeddings. Connecting these two components yields a generative model $P(x|z, text)$ of audio $x$, given caption $z$, and arbitrary content $text$.

\subsection{Emotional VITS}
\label{ssec:evits}
Emotional VITS~(eVITS)\footnote{https://github.com/innnky/emotional-vits} is an open-source TTS model built on VITS \cite{Kin21-CVAW}. It produces emotional speech by directly incorporating emotion embeddings into text embeddings, thereby eliminating the need for emotion labels.
This allows all emotion embeddings to exist in a continuous space, enabling the synthesis of speech with any emotion present in the dataset during inference. 

\subsection{ParaCLAP-NP}
\label{ssec:paraclap}
ParaCLAP is a recently-introduced CLAP-style model specifically designed for Computational Paralinguistics tasks \cite{Jing24-PTA}. It jointly trains parallel audio and text encoders to establish a link between the input audio-text pairs in a shared multimodal space by contrastive learning. We followed the original setup by utilizing a wav2vec 2.0 large model, which is fine-tuned for dimensional Speech Emotion Recognition\footnote{https://huggingface.co/audeering/wav2vec2-large-robust-12-ft-emotion-msp-dim}\cite{WAGNER23-DOT}, as the audio encoder and the BERT  \cite{Devlin18-BPO} base model (uncased) as the text encoder.
Notably, both eVITS and ParaCLAP apply the same audio encoder to extract a reference emotion embedding (during training).
To encourage more direct and consistent integration of emotion embeddings between the two modalities (text and audio), we modified ParaCLAP by removing the projection layer following the audio encoder. We refer to this modified model as ParaCLAP-NP. 
We note that in the Text-to-Image field, the text embedding of CLIP  \cite{Radford21-LTV} can still be transformed into an image by the image decoder with only minimal loss of detail  \cite{Ramesh22-HTI}. 
However, preliminary experiments with ParaCLAP-NP revealed that using its text embeddings directly as emotion embeddings in eVITS led to unintelligible speech generation. This highlights the challenge of aligning text embeddings with emotion embeddings in speech synthesis.

\subsection{Diffusion Model}
\label{ssec:sgms}

\begin{figure}
    \includegraphics[width=0.45\textwidth]{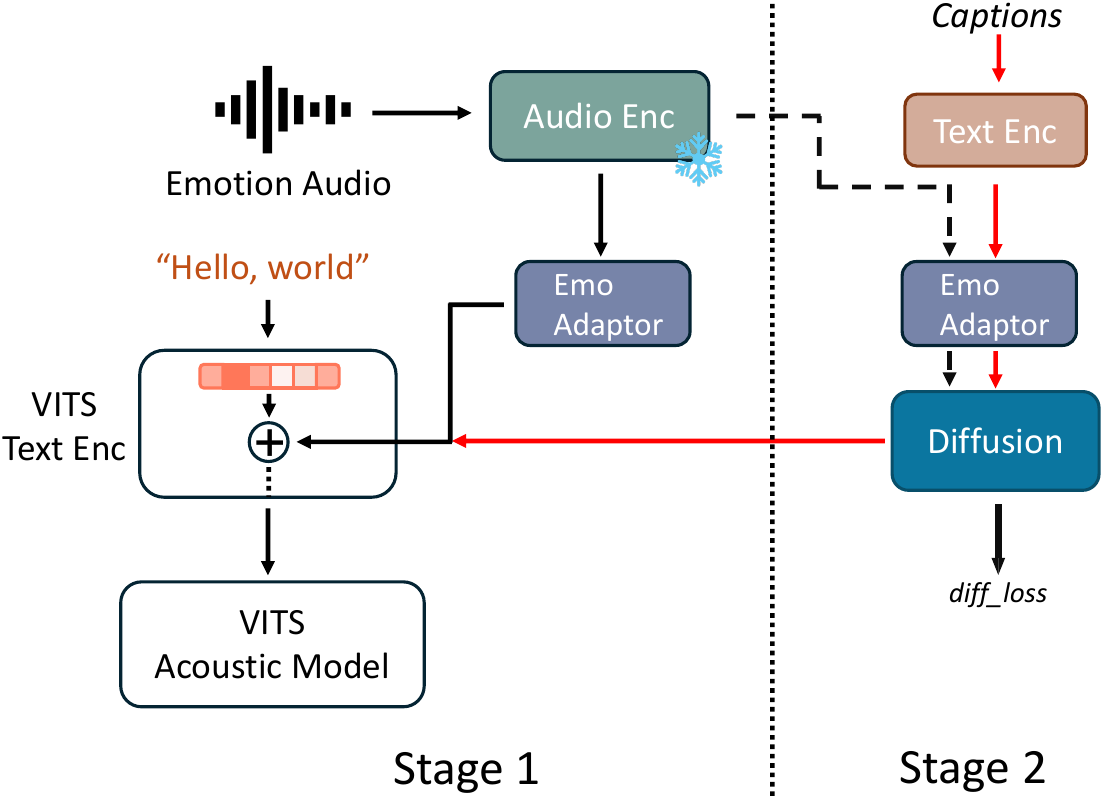}
    \caption{The two-stage training process for our system. Stage 1 involves the training of eVITS using reference acoustic embeddings from an audio encoder. Stage 2 incorporates captioning through a ParaCLAP-NP text encoder followed by a diffusion model. The red lines indicate the components used in the \textbf{inference phase}, where the diffusion model utilizes captions to refine the emotional characteristics in the synthesized speech, and \textbf{no reference audio encoder is used}.}
    \vspace{-4mm}
    \label{fig:arch}
\end{figure}

To address the problem that the text embeddings of ParaCLAP-NP are not directly usable for generating intelligible speech with eVITS, we apply Score-based generative models (SGMs), also known as diffusion models, to bridge the gap between the different modalities. As shown in \cref{fig:arch}, during training, both the emotion embedding from the audio encoder and the caption embeddings from the ParaCLAP-NP text encoder are processed by a pretrained emotion adaptor before being input into the diffusion model. The emo adaptor contains a set of linear layers to map the emotion embedding dimension to that of the eVITS text embedding.

SGMs are proposed by Song et al.\ \cite{Song21-SBG}, who adopt stochastic differential equation~(SDE) formulations \cite{Kloeden92-SDE}, enabling efficient and flexible generation of high-quality samples \cite{Richter23-SEA, Pascual23-FGA}. 
The forward process of SGM$\{x_t\}_t$ can be described by an SDE:

\begin{equation}
\label{equ:sde}
dx_t = f(x_t, t)dt + g(t)dw,
\end{equation}
where $f$ is the $drift$ coefficient of $x(t)$, $g$ is the $diffusion$ coefficient of $x(t)$, and $w$ is a standard Wiener process (or Brownian motion). 

Every SDE in the form of \autoref{equ:sde} should have a corresponding reverse SDE \cite{Song21-SBG}, which is also a diffusion process:

\begin{equation}
\label{equ:rvs_sde}
dx_t= \left[f(x_t, t) - g(t)^2\nabla_{x_t}log p_t(x_t)\right]dt + g(t)d\Bar{w}, 
\end{equation}

where $dt$ is an infinitesimal negative timestep and $\bar{w}$ is a standard Wiener process when $t$ reverses from $T$ to $0$. $\nabla_{x_t}log p_t(x_t)$ is the score function of a prior distribution $p(x)$ which can be approximated by a learned time-dependent score-based model $s_{\theta}$, such that $s_{\theta}(x_t, t) \approx \nabla_{x_t}log p_t(x_t)$.

In this work, we adapted the forward-backward process of the diffusion model from U-Dit TTS \cite{Jing23-UTU} with only one main difference, namely, to account for the emotion embedding as an additional input vector, we modified the 2D UNet in the diffusion model with a 1D UNet. Then, the forward SDE is defined as:
\begin{equation}
\label{equ:fwd}
    dX_t = \frac{1}{2}\Lambda^{-1}(\mu - X_t)\beta_t dt + \sqrt{\beta_t}dW_t, \quad t\in[0, T], 
\end{equation}
where $\beta_t$ is the non-negative-valued noise schedule function, $\mu$ is a vector, and $\Lambda$ is a diagonal matrix with positive elements.

\textbf{Reverse Process:} From \autoref{equ:rvs_sde} and the simplified reverse procedure from the probability flow ordinary differential equation~(ODE) \cite{Song21-SGM}, we have our reverse process:

\begin{equation}
    \label{equ:rev_sde}
    dX_t=\left(\frac{1}{2}\Lambda^{-1}(\mu- X_t) - \nabla logp_t(X_t)\right)\beta_tdt.
\end{equation}

\begin{figure*}[t]
    \centering
    \vspace{-5mm}
    \includegraphics[width=0.96\textwidth]{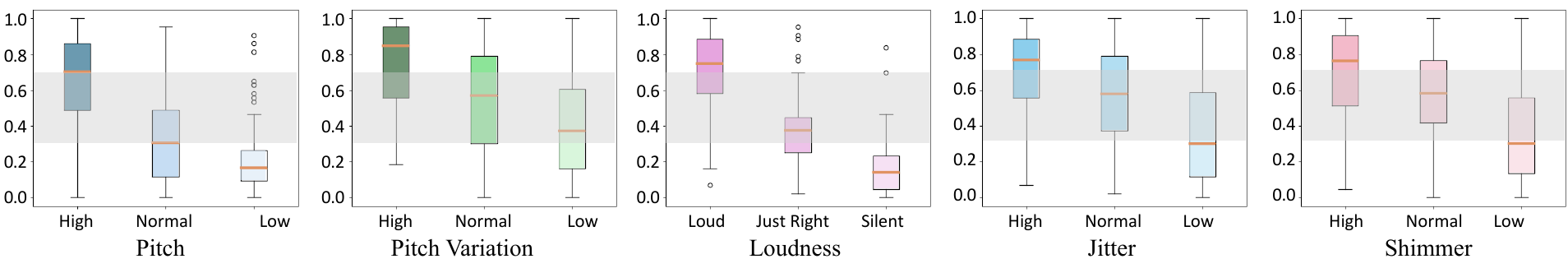}
    \caption{Speech attribute controllability. The x-axis represents the intensity descriptions of various attributes in the captions, while the y-axis indicates the numerical ranges of these attributes. The gray area in the figure divides the values into three parts, representing \textbf{low/normal/high} intensities from bottom to top, respectively.}
    \label{fig:var}
\end{figure*}

\subsection{Training \& Inference}
\noindent
\textbf{Training -- Stage I:} In the first stage of training, we train eVITS using reference audio processed by the audio encoder and the emotion adaptor. This serves to bootstrap training and initialize the TTS model and adaptation modules.

\noindent
\textbf{Training -- Stage II:} During the second training stage, text and audio embeddings are derived from matching caption-reference pairs. They are processed by the pretrained emotion adaptor from Stage I and then fed to the diffusion model in parallel. This setup allows the model to accommodate and integrate both sources of information effectively.

\noindent
\textbf{Inference:} During inference, the trained eVITS model is conditioned \emph{solely on text captions} -- processed by the emotion adapter and diffusion model -- without using reference audio.

\section{Experiments}
\label{sec:exp}
\subsection{Dataset}
\label{ssec:dataset}
%for paraclap training
For ParaCLAP-NP training, we used MSP-Podcast Release 1.9 \cite{Lotfian17-BNE}, which consists of natural English speech extracted from podcast recordings. The dataset includes 55,283 utterances from over 1,200 speakers, spanning more than 110 hours of speech. We adapted the query generation pipeline from ParaCLAP to construct the text-audio pairs~ \cite{Jing24-PTA}. Details are presented in \cref{ssec:query}.

%for evits training
For eVITS training, we utilized the English portion of the Emotional Speech Database~(ESD)  \cite{Zhou21-SAU}. It contains 350 parallel utterances from 10 native English speakers, covering 5 emotion categories: \textbf{neutral, happy, angry, sad, and surprise}. We set a single-speaker training on data of speaker 0016, a highly-expressive female.

\subsection{Query Generation}
\label{ssec:query}
We generate text queries\footnote{The process can be more easily understood by inspecting the code submitted as supplementary material.} matching each sample by relying on two main sources of information:
a) The labels already included in the dataset; 
b) Expert acoustic and prosodic features extracted for each sample.
Each source of query is extended to a complete sentence by the template as described below.

\noindent
\textbf{Dataset labels:}
MSP-Podcast is labeled for emotion and gender. We fix the emotion queries by ``speaker is [EMOTION]", with [EMOTION] coming in the form of an adjective (\eg ``angry'', ``happy'', etc.).

\noindent
\textbf{Pseudo-captions using expert features:} The
Extended Geneva Minimalistic Acoustic Parameter Set (eGeMAPS)~ \cite{Eyben15-TGM} is extracted by the openSMILE toolkit \cite{Eyben10-OTM}. 
In this study, we focus on the following features which are easier to interpret:
a) mean ($\mu$) and standard deviation ($\sigma$) of \emph{pitch};
b) sound \emph{intensity} (over the whole utterance);
c) \emph{jitter} (mean over utterance);
d) \emph{shimmer} (mean over utterance).
This gives us 4 numerical features, which we proceed to bin according to their distribution in the training set; we do the same for the dimensional emotional attributes (arousal, valence, dominance).
Following that, we generate queries using those bins (\eg ``pitch is low/normal/high'').

\subsection{Experiment Settings}
\begin{itemize}
    \item \textbf{ParaCLAP-NP:} We employed the Adam optimizer with a batch size of 64 and set the number of training epochs to 50. The text branch was fine-tuned with a learning rate of $1e-5$. Meanwhile, all other parameters employed a learning rate of $1e-3$. The model from the epoch that yielded the best performance on the MSP-Podcast test set is selected for the following task;
    \item \textbf{eVITS:} we used an Adam optimizer with learning rate $2e-4$ and a batch size of 128. All the training audios were resampled to 22.05 kHz;
\end{itemize}
All of our models were trained and tested in a Python 3.10.8 and PyTorch 1.13.1 environment.

\section{Results and Analysis}
To validate the performance of our system, we designed and conducted a series of experiments, including both subjective and objective evaluations. In subjective experiments, we invited 20 participants, comprising 11 males and 9 females, to assess the quality and emotional expressiveness of the synthesized speech. Participants were instructed to use headphones and rate the samples on a predefined scale, ensuring consistency in the evaluation process.

\label{sec:res}
\subsection{Speech Quality}
To assess the speech quality of our proposed system, we conducted a Quality Mean Opinion Score~(MOS-Q) experiment. Participants were asked to rate the naturalness of speech samples, uniformly selected across different emotions, on a 5-point scale, where 1 represents ``very bad'' and 5 represents ``excellent''.

Except for our proposed system, there are other three different systems involved in the MOS-Q experiment:
\begin{itemize}
    \item \textit{Ground Truth}: the original audio sample from the ESD dataset;
    \item MixedEmotion \cite{Zhou23-SSW}:  An autoregressive~(AR) emotional TTS system that leverages precomputed emotion intensity values for controllability and allows manual adjustments at run time;
    \item eVITS: The backbone of ParaEVITS, which can generate emotional audio using only the corresponding emotion embeddings.
\end{itemize} %

\cref{tab:q-mos} presents the MOS-Q experiment results with a 95\% confidence interval~(CI) for these four systems. Among the models tested, \evits shows the best performance on both the MOS score and CI value.
The narrow margin between the CI value of the  \textit{Ground Truth} and \evits indicates a high level of agreement among participants. 
However, the gap between our system and the Ground Truth indicates that there is still room for improvement in quality. This will be a focus for further study in our future work.

\subsection{Speech Expressiveness}
Benefiting from the generative capability of diffusion models, \evits can produce diverse samples within the same emotional category. We also conducted a subjective Emotion Similarity MOS Experiment (MOS-S) to measure the emotional similarity between different speech samples generated by our system. In this experiment, participants 
were required to listen to pairs of speech samples that were generated from identical prompts but different initial states. This simulates the use-case often seen in generative models (e.\,g., DALL-E) that produces multiple `candidate' outputs for a given prompt and lets the user select the most suitable one. 
Participants rated the emotional similarity of each pair on a 5-point scale, where 1 indicates completely different emotional styles~(worst), and 5 indicates that the samples have identical emotional styles with distinct expressions~(best).

\begin{figure}
    \centering
    \includegraphics[width=0.42\textwidth]{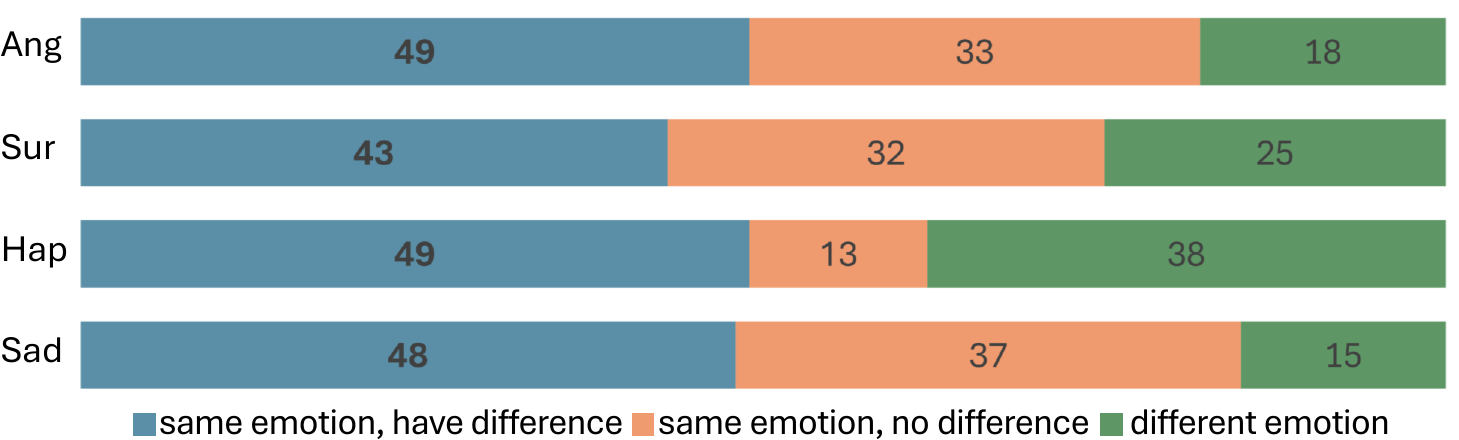}
    \caption{Emotion Similarity Preference Test Results}
    \vspace{-3mm}
    \label{fig:s-mos}
\end{figure}

Additionally, we categorized the score range into three distinctive classes: $score < 3$ indicates that the two samples contain different emotions, $score=3$ shows that the two samples have identical emotions but with minor difference in style, and $score > 3$ means that the two samples have the same emotion but with sufficient style difference. The results of the MOS-S is presented in \cref{tab:s-mos} and  \cref{fig:s-mos} demonstrates emotion similarity preference result.

%BS: Please add an upwards poiting arrow afer MOS-Q (also inthe other table after MOS-S) to show that higher is better and try to align the numbers 
\begin{table}
\centering
    \caption{Quality Mean Opinion Score Experiment (MOS-Q) with 95\% Confidence Interval}
    \label{tab:q-mos}
    \begin{tabular}{l | r } 
    \toprule
    Model & MOS-Q\,$\uparrow$\\ 
    \midrule
     \textit{Ground Truth} & $4.40 \pm\, 0.082$ \\ 
     \midrule
     MixedEmotion (baseline) & $2.23 \pm\, 0.097$\\ 
    
    eVITS & $3.70 \pm\, 0.106$ \\ 
    
    \evits & $\mathbf{3.88} \pm\,\mathbf{0.095}$\\
    \bottomrule
    \end{tabular}
    \vspace{-4mm}
\end{table}

As shown in \cref{fig:s-mos}, on every emotion category, approximately half of the participants evaluated the test audio pairs as having the same emotion but with noticeable differences in expression. Moreover, in \cref{tab:s-mos} we observe that the scores for Angry and Sad samples are very close. However, the Angry samples have a lower standard deviation, indicating that the emotion of anger is generated with better consistency. The results on happy exhibited the lowest MOS-S scores in the experiment, showing an overall worse outcome, with differences in initialization manifesting as differences in perceived emotion, rather than style. As depicted in \cref{fig:s-mos}, while 49\% of the participants perceived clear differences within the happy samples, over 38\% of the participants believed the test audios represented entirely different emotions. 
Broadly, these results show that emotional TTS models may not perform equally well across all emotions -- ours, in particular, struggles most for ``happy'' speech as compared with other emotions, a fact warranting a deeper investigation in future work.
Despite that, results show ParaEVITS's capability to generate diverse emotional speech.

\subsection{Speech Attribute Controllability}
As mentioned in \cref{sec:intro}, a caption typically includes five audio attributes besides the emotion label to control the final output. To further investigate the impact of different audio attributes on the generated results, we conducted an objective experiment. We selected 44 captions covering all attribute variations, and generated 20 audio samples per caption under different initial states and transcriptions.
\textbf{Crucially, we manipulated the intensity of each parameter with an adjective quantifier (high, normal, low for all except loudness, where we used `loud', `silent', and `just right' to evaluate the ability of the model to capture diverse linguistic queries)}.
All the generated speech was then processed by openSMILE to extract the same eGeMAPS features that were used to create the captions.

\begin{table}
\centering
    \caption{Emotion Similarity Mean Opinion Score Experiment (MOS-S) with 95\% Confidence Interval}
    \label{tab:s-mos}
    \begin{tabular}{l | r } 
    \toprule
    Emotion & MOS-S\,$\uparrow$\\ 
     \midrule
     Angry & $3.36 \pm\, 0.187$ \\ 
    
     Surprise & $3.22 \pm\, 0.268$\\ 
    
    Happy & $3.05 \pm\, 0.234$ \\ 
    
    Sad & $3.34 \pm\,0.230$\\
\bottomrule
\end{tabular}
\end{table}

\cref{fig:var} provides a detailed overview of the performance across various audio attributes.
It can be observed that while there is overlap in the results of each attribute's variations, there are still clear differences among the statistical outcomes.
Moreover, each attribute displays distinct patterns in terms of central tendency, variability, and consistency.
`High levels' of Pitch, Pitch Variation, Jitter, and Shimmer exhibit considerable variability, 
indicating a diversity in system outputs that does not always conform to specifications (with results oftentimes being in the `normal' range).
`Normal levels' show less variability, suggesting better consistency and controllability.
`Low levels' of Pitch and Loudness are the most consistent overall, indicating higher specificity when controlling these two attributes.
In contrast, Loudness shows higher consistency in its `Just Right' (\ie `normal') range, while the Loud and Silent categories show more variability and overlap with the normal range. 
Overall, these results indicate that our system is capable of capturing and generating diverse audio attributes \emph{controllable via rich textual prompts}, with certain settings demonstrating more consistency than others.

\section{Conclusion}
\label{sec:con}
In this work, we introduced ParaEVITS, a ParaCLAP-based natural language compositionality to enhance the expressive of emotional TTS systems. We conditioned a diffusion model to control paralinguistic attributes with text prompts. This Computational Paralinguistics-based approach bridged low-level acoustic features and high-level emotion descriptions, enabling rich emotional control for synthesized speech. The diffusion process generated effective natural language emotion embeddings, guiding the TTS framework to produce emotion speech that aligns with natural language prompts without compromising speech fidelity. 
Future works include multispeaker support functionality, LLM-based captions generation, and applying it to more challenging and diverse datasets, such as conversional speech.

\section{Acknowledgment}
Björn W.\ Schuller is also with the Munich Data Science Institute (MDSI) and the Munich Center for Machie Learning (MCML), all in Munich, Germany. We are grateful to the China Scholarship Council~(CSC), Grant \#\,202006290013, and the DFG (German Research Foundation),  Reinhart Koselleck-Project AUDI0NOMOUS (Grant No.\ 442218748) to support this work. We would also like to thank Prof. Jiangjian Xie and all the people involved in the MOS experiments for their invaluable contribution to this work.
\newpage
\clearpage
\bibliographystyle{IEEEtran}
\bibliography{sample}

\end{document}